\documentclass[sigconf, nonacm]{acmart}
\usepackage[linesnumbered, ruled, vlined]{algorithm2e}
\usepackage{algpseudocode}
\usepackage{amsmath}

\usepackage{amssymb}
\usepackage{graphicx}
\usepackage{subcaption}
\usepackage{tabularx}
\usepackage{multirow}
\usepackage{diagbox}
\usepackage{natbib}
\usepackage{bm} 
\usepackage[most]{tcolorbox}
\usepackage{graphicx} 
\usepackage{caption} 
\usepackage{booktabs}
\usepackage{array}
\usepackage{pifont}
\newcommand{\cmark}{\ding{51}}  
\newcommand{\xmark}{\ding{55}}  
\usepackage{siunitx}

\AtBeginDocument{%
  }

\title{LLMVul: A Vulnerability-Labeled Dataset of LLM-Generated C/C++ Functions from Real Production Repositories}

\author{Mohammad Farhad} 
\affiliation{%
\institution{School of Computing \& Informatics \\University of Louisiana at Lafayette} \city{Lafayette} \state{LA} \country{USA} } \email{mohammad.farhad1@louisiana.edu} \author{Shuvalaxmi Dass} 
\affiliation{%
\institution{School of Computing \& Informatics \\University of Louisiana at Lafayette} \city{Lafayette} \state{LA} \country{USA} } \email{shuvalaxmi.dass@louisiana.edu}

\renewcommand{\shortauthors}{Mohammad et al.}

\begin{abstract}
Large language models (LLMs) are increasingly used to generate and assist with software development, yet existing vulnerability datasets largely focus on human-written code or controlled prompting environments. This limits the ability to study security weaknesses in LLM-generated code as it appears in real-world software projects. We present LLMVul, a vulnerability-labeled dataset of LLM-generated C/C++ functions mined from real production repositories. We mine AI-assisted development activity from GitHub over a 4 year period, from November 13, 2022 to September 3, 2026, using provenance signals such as commit metadata and AI-related authorship evidence. After filtering and deduplication, LLMVul contains 21,430 unique C/C++ functions from 226 repositories, together with repository, commit, function, provenance, and AI-tool metadata. We establish vulnerability labels using an ensemble of complementary static-analysis and pattern-based techniques and assign Common Weakness Enumeration (CWE) categories to confirmed vulnerable functions. To assess labeling reliability, we additionally conduct independent manual annotation and measure inter-rater agreement using Cohen's kappa ($k=0.79$). LLMVul contains 1,540 ensemble-vulnerable functions spanning 17 unique CWE categories, providing substantially more real-world LLM-generated vulnerable C/C++ functions than existing vulnerability-oriented LLM code benchmarks. By preserving both code-level vulnerability labels and generation/provenance metadata, LLMVul enables reproducible research on vulnerability detection, security evaluation of LLM-generated code, and analysis of vulnerability patterns in AI-assisted software development. The LLMVul dataset is publicly available at \url{https://doi.org/10.5281/zenodo.22668216}.
\end{abstract}

\ccsdesc[500]{Security and privacy}
\ccsdesc[500]{Security and privacy~Software security and testing}
\ccsdesc[500]{Security and privacy~Vulnerability analysis}
\ccsdesc[500]{Security and privacy~Software engineering}

\begin{document}

\keywords{Source Code Analysis, Vulnerability Detection, LLM-generated Code, Dataset, Software Security, Mining Software Repositories}

\maketitle
\section{INTRODUCTION}
\label{sec:introduction}

The widespread adoption of large language model (LLM)-based coding assistants has fundamentally altered how software is produced. GitHub Copilot surpassed one million active users within months of its general availability in June 2022 \cite{copilot2022ga}, and by 2024 reported over 1.8 million paid subscribers \cite{copilot2024subscribers}.
Tools such as ChatGPT, Claude, Cursor, and Google Gemini are now routinely used to generate
production code across security-sensitive domains including systems software, embedded firmware, and network services \cite{wang2025aicode}. Recent large-scale measurements confirm that approximately 20--30\% of new commits in major open-source repositories contain AI-attributed code \cite{wang2025aicode,liu2026debtaiboom}, with projections suggesting this proportion will continue to grow \cite{amodei2025machines}.

This shift introduces a security risk that the research community has only begun to examine. LLM-generated code is known to exhibit elevated vulnerability rates: Siddiq and Santos \cite{siddiq2022securityeval} found that 74\% of Copilot-generated Python functions contained security weaknesses detectable by human experts, while Perry et al. \cite{perry2023security} showed that developers who used AI assistance were \emph{significantly more likely} to produce insecure code than those who did not. A 2026 formal verification study of 3,500 LLM-generated artifacts reported that six industry-standard static analysis tools combined detected only 7.6\% of formally proven vulnerabilities, missing 97.8\% of exploitable code \cite{blain2026broken}. Most recently, SecRepoBench demonstrated that GPT-5---the most capable model currently available---achieves only 39.3\% secure-pass@1 on real-repository code completion tasks \cite{secrepobench2025}, confirming that the security problem persists even with the best available models and agent frameworks.

\indent\textbf{The dataset gap.} Despite this growing body of evidence, the datasets available to the research community for studying LLM-generated code security share a critical structural limitation: \emph{existing benchmark mostly rely on controlled, researcher-directed prompts rather than from code that developers actually committed to production repositories}. SecurityEval \cite{siddiq2022securityeval} provides 130 hand-crafted prompts designed to elicit specific CWE patterns from code generation models.
CyberSecEval 2 \cite{bhatt2024cyberseceval2} and SafeGenBench \cite{safegenbench2025} follow the same paradigm at larger scale---558 and approximately 500 samples respectively---while CWEval \cite{cweval25} deliberately isolates functions from third-party dependencies to simplify evaluation. As a consequence, existing LLM security benchmarks primarily measure how often LLMs generate specific vulnerabilities under explicit security-oriented prompts, capturing a controlled laboratory phenomenon rather than real-world developer behavior. In production, however, security vulnerabilities may emerge incidentally when developers use LLMs for routine feature implementation without any explicit security framing. This gap leaves an important question largely unanswered: \textit{How frequently do LLMs introduce security vulnerabilities during ordinary, non-security-focused code generation?}

Two recent empirical studies have examined the security of LLM-generated code in real-world settings, yet neither provides labeled vulnerability data for systematic security analysis. Wang et al. \cite{wang2025aicode} mine AI-attributed commits from 1,000 GitHub repositories and confirm that certain CWE families are overrepresented in AI-tagged commits---but provide no \textit{function-level} vulnerability labels suitable for training or evaluating detection models. Liu et al. \cite{liu2026debtaiboom} analyse technical debt in 3,02,600 AI-authored commits across 6,299 repositories but focus on Python, JavaScript, and TypeScript rather than the C/C++ ecosystem, and similarly provide no vulnerability labels. On the human-written side, datasets such as BigVul \cite{Bigvul}, Devign \cite{Devign}, and PrimeVul \cite{primeVul} provide high-quality labeled C/C++ vulnerability corpora, but are composed entirely of human-authored code. Recent work has demonstrated that models trained on these corpora may not generalise to LLM-generated
code, whose vulnerability patterns differ structurally from human-written code \cite{primeVul}.

\indent\textbf{Our work.} We present \textbf{LLMVul}, the first vulnerability-labeled dataset of C/C++ functions mined from \emph{real production repositories} (as projects shown in Table \ref{tab:top_repos_functions}) where code was generated by AI coding assistants. LLMVul bridges the gap between laboratory benchmarks and real-world deployment: rather than prompting an LLM to produce vulnerable code, we identify functions that developers committed to GitHub after using AI coding assistants, including GitHub Copilot, ChatGPT, Claude Code, and Cursor, and assess their security using an ensemble of three static-analysis tools---Semgrep, Flawfinder, and pattern-matching. The resulting vulnerability labels were subsequently validated on a representative sample by human raters, achieving substantial inter-rater agreement ($\kappa = 0.79$). The dataset comprises 21.4k unique C/C++ functions drawn from \num{226} repositories and 1,684 unique LLM-attributed commits spanning November 2022 to September 2026, of which 1,540 (7.2\%) are labeled vulnerable across 17 unique CWE categories. Each function is annotated with full commit provenance, AI tool attribution, individual tool verdicts, and CWE identifiers, enabling a range of research tasks described in Section \ref{sec:research-questions}.

\section{RELATED WORK}
Existing research on the security of AI-generated code encompasses three broad categories of datasets: human-written vulnerability benchmarks, LLM-prompted code datasets, and datasets mined from AI-assisted software development in public repositories. Building on this taxonomy, Table \ref{tab:dataset_comparison} compares LLMVul with representative existing benchmarks across key dimensions, highlighting differences in data provenance, code-generation context, vulnerability coverage, and labeling methodology.

\section{DATASET CURATION}
\label{sec:curation}

LLMVul is constructed through a three-phase pipeline: (1) mining LLM-attributed C/C++ commits from public GitHub repositories, (2) extracting function-level code units, and (3) labeling extracted functions using a three-tool static analysis ensemble validated by human raters. Table \ref{tab:dataset_features} summarizes all 34 features recorded per function.

\subsection{Phase 1: Repository Mining and AI Attribution}
\label{subsec:mining}
\noindent\textbf{Repository selection.} We queried the GitHub REST API for public C and C++
repositories with at least 200 stars and a commit activity after June 2022—the general availability date of GitHub Copilot \cite{copilot2022ga}. In total we scanned 1,200 repositories and 321k commits, identifying 7,018 LLM-attributed commits from 226 repositories containing C/C++ functions (Table \ref{tab:top_repos_functions} showing top 10 repositories).

\noindent\textbf{AI attribution signals.} A commit is classified as \emph{LLM-attributed} if its message or metadata contains at least one of two categories of signal, following the attribution methodology of Liu et al.~\cite{liu2026debtaiboom}:

\begin{itemize}
    \item \textbf{Strong signals} (\texttt{signal\_strength} $=$ \texttt{strong}): explicit co-authorship tags injected by the AI tool itself into the Git trailer, e.g.\ \texttt{Co-authored-by:\ GitHub Copilot}, \texttt{Co-authored-by:\ cursor-noreply}, or \texttt{Co-authored-by:\ claude-code}. These tags are machine-generated and carry the highest attribution confidence.
    \item \textbf{Medium signals} (\texttt{signal\_strength} $=$ \texttt{medium}): developer authored phrases in the commit message, e.g.\ \emph{``generated with ChatGPT''}, \emph{''copilot suggested''}, \emph{''ai-assisted''} or \emph{''via cursor AI''}. These phrases are self-declared by the developer and are slightly less certain than co-authorship tags.
\end{itemize}

\begin{table}[t]
\centering
\footnotesize
\caption{Top 10 repositories by number of extracted functions.}
\label{tab:top_repos_functions}
\begin{tabular}{clc}
\hline
\textbf{Rank} & \textbf{Repository} & \textbf{Functions} \\
\hline
1  & Serial-Studio/Serial-Studio     & \textbf{6,159} \\
2  & nature-lang/nature              & \textbf{1,132} \\
3  & microsoft/ebpf-for-windows      & \textbf{699} \\
4  & isl-org/Open3D                  & \textbf{466} \\
5  & microsoft/onnxruntime           & \textbf{465} \\
6  & DarkFlippers/unleashed-firmware & \textbf{460} \\
7  & DavidXanatos/TaskExplorer       & \textbf{452} \\
8  & carla-simulator/carla           & \textbf{446} \\
9  & microsoft/WSL                   & \textbf{410} \\
10 & sqliteai/warp                   & \textbf{378} \\
\hline
\end{tabular}
\end{table}

\begin{figure}[htbp]
    \centering
    \includegraphics[width=0.25\textwidth]{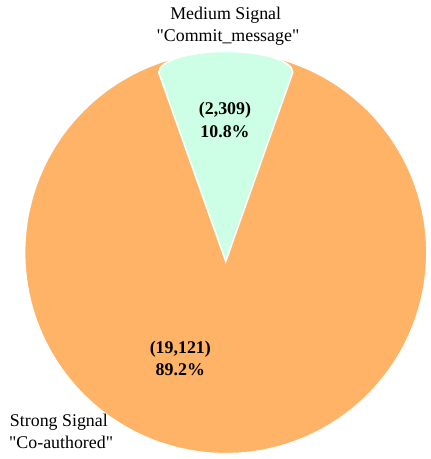} 
    \caption{Signal strength identified by the pipeline.}
    \label{fig:signal-strength}
\end{figure}

\noindent The pipeline identified 89.2\% strong-signal and 10.8\% medium-signal attributed functions, as illustrated in Figure \ref{fig:signal-strength}. These signals were derived from textual indicators of AI-assisted code generation. In total, we compiled up to \num{47} regular-expression patterns to detect these indicators across nine AI coding assistants: GitHub Copilot, ChatGPT, Claude / Claude Code, Cursor, Gemini, Devin, and OpenAI Codex. Claude Code accounts for the largest share of extracted functions, followed by GitHub Copilot (see Table \ref{tab:ai_tool_breakdown}). Of the 321k commits scanned, 7,018 (2.2\%) matched at least one attribution pattern, a proportion consistent with Wang et al. \cite{wang2025aicode}, who report approximately 2\% AI-attributed commits in production repositories. Attribution metadata is preserved in the \texttt{ai\_tool}, \texttt{signal\_strength}, and \texttt{signal\_type} columns.

\begin{table}[t]
\centering
\footnotesize
\caption{Distribution of extracted functions identified by AI coding tool.}
\label{tab:ai_tool_breakdown}
\begin{tabular}{lc}
\hline
\textbf{AI Tool (Total 9)} & \textbf{Functions} \\
\hline
Claude Code  & 14,221 \\
GitHub Copilot & 4,952 \\
Claude & 807 \\
Gemini & 386 \\
Cursor & 373 \\
ChatGPT & 298 \\
Generic LLM & 242 \\
OpenAI Codex & 135 \\
Devin & 16 \\
\hline
\end{tabular}
\end{table}

\begin{table}[t]
\centering
\footnotesize
\caption{Summary of the mined C and C++ functions including repository statistics, commit metrics, lines of code (LoC) constraints and language distribution.}
\label{tab:dataset_overview}
\begin{tabular}{lr}
\hline
\textbf{Metric} & \textbf{Value} \\
\hline
Total functions & \textbf{21,430} \\
Repositories Mined & \textbf{1,200} \\
LLM-attributed repositories & \textbf{226} \\
Total commits scanned & \textbf{3,21,080} \\
LLM commits found & \textbf{7,018} \\
Unique commits & \textbf{1,684} \\
Date range & Nov 13, 2022 --- Sep 3, 2026 (\textbf{4 Years}) \\
Maximum LoC & \textbf{150}\\
Minimum LoC & \textbf{5}\\
\hline
\hline
\textbf{Language} & \textbf{Functions} \\
\hline
\textbf{C++} & 12,546 (\textbf{58.5\%}) \\
\textbf{C} & 8,884 (\textbf{41.5\%})\\
\hline
\end{tabular}
\end{table}

\subsection{Phase 2: Function Extraction}
\label{subsec:extraction}

\begin{figure}[htbp]
    \includegraphics[width=7.5cm, height=6cm]{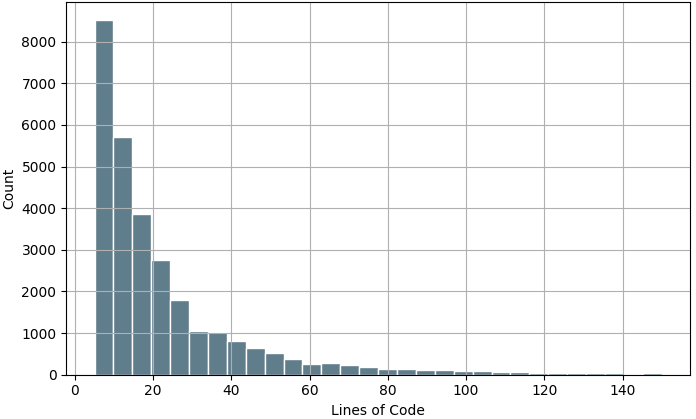} 
    \caption{Lines of code (LoC) distribution in the dataset.}
    \label{fig:funnction-length}
\end{figure}

For each LLM-attributed commit we retrieved the file-level differences via the GitHub API and extracted only the \emph{added or modified} lines (i.e.\ lines prefixed with \texttt{+} in the unified diff format). Function boundaries were identified using the \emph{tree-sitter} incremental parser~\cite{treesitter} with its C and C++ grammars, targeting \texttt{function\_definition} AST nodes. We applied two filters: functions shorter than five lines or longer than 150 lines were discarded as either too trivial or too large for reliable function-level analysis. Figure \ref{fig:funnction-length} illustrates the distribution of function lengths, measured in lines of code (LoC), for the functions retained in the final dataset. Each retained function is stored with its source coordinates (\texttt{function\_start\_line}, \texttt{function\_end\_line}, \texttt{function\_lines}), the containing file (\texttt{file\_name}, \texttt{file\_hash}), and full commit provenance (\texttt{commit\_id}, \texttt{commit\_url}, \texttt{commit\_message}, \texttt{commit\_date}). A 16-character hexadecimal \texttt{unique\_id} ties each record to its exact extraction context.

After extraction and deduplication removing identical function bodies that appeared in multiple commits--21,430 unique functions remained. A detailed breakdown is provided in Table \ref{tab:dataset_overview}.

\begin{table*}[htbp]
\centering
\caption{Features and metadata contained in the LLMVul dataset.}
\label{tab:dataset_features}
\footnotesize
\begin{tabular}{p{2.8cm}p{4.0cm}p{6.2cm}}
\hline
\textbf{Feature Category} & \textbf{Column Name} & \textbf{Description} \\
\hline
Function Id & \texttt{unique\_id} & Unique identifier for each function. \\
Project Name & \texttt{project\_name} & GitHub repository containing the function. \\
Project URL & \texttt{project\_url} & URL of the GitHub repository. \\
Commit Id & \texttt{commit\_id} & Git commit identifier from which the function was extracted. \\
Commit URL & \texttt{commit\_url} & URL of the corresponding commit. \\
Commit Message & \texttt{commit\_message} & Commit message associated with the function. \\
Commit Date & \texttt{commit\_date} & Date of the corresponding commit. \\
File Name & \texttt{file\_name} & Source file containing the function. \\
File Hash & \texttt{file\_hash} & Hash of the source file. \\
Source Code Language & \texttt{language} & Programming language (C/C++). \\
Function Information & \texttt{function\_body} & Extracted source code of the function. \\
Function Startline & \texttt{function\_start\_line} & Starting line of the function. \\
Function Endline & \texttt{function\_end\_line} & Ending line of the function. \\
Function Length & \texttt{function\_lines} & Number of lines in the function. \\
AI Provenance & \texttt{ai\_tool} & AI coding assistant associated with the code. \\
AI Signal-Strength & \texttt{signal\_strength} & Strength of evidence for AI-generated code. \\
AI Signal Type & \texttt{signal\_type} & Type of evidence identifying AI-generated code. \\
Repository Stars & \texttt{repo\_stars} & Number of repository stars. \\
Repository Forks & \texttt{repo\_forks} & Number of repository forks. \\
Project Primary Lang. & \texttt{repo\_primary\_language} & Repository's primary language. \\
Dataset Timestamp & \texttt{mined\_at} & Timestamp when the function was collected. \\
Vulnerable Classification & \texttt{vuln\_label} & Final vulnerability classification. \\
Ensemble Label & \texttt{ensemble\_label} & Result of the ensemble analysis. \\
Semgrep Tool Results & \texttt{tool\_semgrep} & Whether Semgrep flagged the function. \\
Flawfinder Tools Results & \texttt{tool\_flawfinder} & Whether Flawfinder flagged the function. \\
Pattern Matching Results & \texttt{pattern\_match} & Whether pattern matching flagged the function. \\
CWE Id & \texttt{cwe\_id} & Final CWE assigned to the function. \\
CWE Information & \texttt{cwe\_description} & Description of the final CWE. \\
CWE URL & \texttt{cwe\_url} & CWE URL when a corresponding CVE is available. \\
Semgrep-Level CWE & \texttt{cwe\_semgrep\_raw} & CWE reported by Semgrep. \\
Flawfinder-Level CWE & \texttt{cwe\_flawfinder\_raw} & CWE obtained from Flawfinder. \\
Pattern-Level CWE & \texttt{cwe\_pattern-match\_raw} & CWE obtained from pattern matching. \\
Semgrep Metadata & \texttt{semgrep\_rule} & Semgrep rule that triggered the finding. \\
Flawfinder Metadata & \texttt{flawfinder\_risk} & Flawfinder risk level from 0 (lowest) to 5 (highest). \\
\hline
\end{tabular}
\end{table*}

\subsection{Phase 3: Vulnerability Labeling}
\label{subsec:labeling}

\noindent\textbf{Three-tool ensemble.} Each function was independently analysed by three complementary static analysis tools operating directly on the extracted function body:

\indent \textbf{1. Semgrep} \cite{semgrep}: a pattern-matching SAST engine applied with CWE-based rules targeting dangerous function calls, format string sinks, and command injection patterns. Results are stored in \texttt{tool\_semgrep} and \texttt{cwe\_semgrep\_raw}.

\indent \textbf{2. Flawfinder} \cite{flawfinder}: a lexical scanner that identifies calls to functions in its built-in risk database, assigning a risk level from 1 (lowest) to 5 (highest). Only findings at level~1 or above are retained. Results are stored in \texttt{tool\_flawfinder}, \texttt{cwe\_flawfinder\_raw}, and \texttt{flawfinder\_risk}.

\indent \textbf{3. Pattern matching}: a curated set of 54 regular-expression patterns informed by the NIST Software Assurance Reference Dataset (SARD) dangerous-function taxonomy and the ITS4 vulnerability scanner \cite{its4}. Each pattern maps directly to a CWE identifier. Results are stored in \texttt{pattern\_match} and \texttt{cwe\_pattern\_match\_raw}.

\smallskip
\noindent\textbf{Majority-vote ensemble.} A function is assigned \texttt{vuln\_label}~$= 1$
(\emph{vulnerable}) if at least two of the three tools flag it; \texttt{vuln\_label}~$= 0$ (\emph{safe}) if none flags it; and \texttt{ensemble\_label}~$=$ \texttt{uncertain} if only one tool flags it. Uncertain functions are excluded from the labeled training set but retained in the dataset for exploratory analysis. The final CWE identifier (\texttt{cwe\_id}) is assigned by priority --- Semgrep $\succ$ Flawfinder $\succ$ pattern matching -- preferring to prioritize the most specific tool-supported CWE over the generic CWE-676 fallback. CWE descriptions and CWE reference URLs are appended from the MITRE CWE catalogue~\cite{mitre_cwe} and stored in \texttt{cwe\_description} and \texttt{cwe\_url}.

\noindent\textbf{Manual validation and inter-rater agreement.} To assess label reliability, two independent raters (the authors) manually inspected a stratified random sample of 100 ensemble-vulnerable functions, drawn proportionally across CWE categories. Each rater independently assigned a binary label (1 = vulnerable, 0 = false positive) based solely on the function body and the assigned CWE identifier, without knowledge of the other rater's labels or the tool outputs. Inter-rater agreement yielded Cohen's $\kappa = 0.79$, indicating \emph{substantial agreement} \cite{landis1977}. Disagreements were resolved through discussion and consensus; the consensus labels supersede the ensemble labels for the 100 validated functions.

\begin{figure}[htbp]
    \centering
    \includegraphics[width=0.5\textwidth]{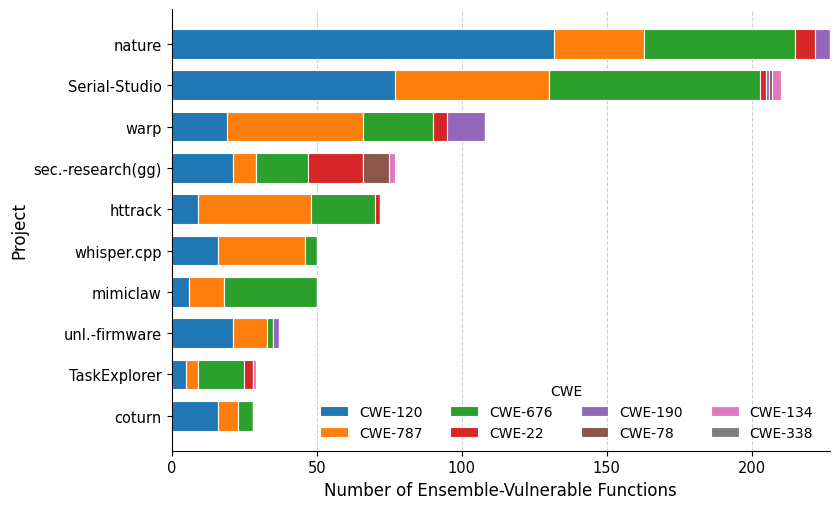} 
    \caption{CWE distribution across the top 10 projects with the highest number of ensemble-vulnerable functions.}
    \label{fig:cwe-visualization}
\end{figure}

\noindent\textbf{Label distribution.}
Table \ref{tab:vulnerability_summary} summarizes the outcome of the labeling process.
Of the 21,430 functions, 1,540 (7.2\%) are labeled vulnerable across eight CWE categories by the three-tool ensemble tool, with CWE-120 (Buffer Copy) and CWE-787 (Out-of-Bounds Write) being the most prevalent (see Table \ref{tab:top_cwe}), and 17,211 (80.3\%) are labeled safe, and 2,679 (12.5\%) remain uncertain hence excluded from the binary-labeled set. The 7.2\% observed vulnerability rate preserves the naturally occurring class distribution in our real-world sample, rather than artificially balancing vulnerable and non-vulnerable instances. Importantly, Flawfinder flagged most of the vulnerable functions as \textit{dangerous function use}, that could not be precisely mapped to a more specific CWE. These cases were subsequently verified under the generic CWE-676 category, as shown in Table \ref{tab:flawfinder_categories}, with buffer-related issues constituting 84.6\% of all classified findings. Table \ref{tab:top_projects_cwe} reports the top ten projects with the highest concentration of vulnerable functions. Figure \ref{fig:cwe-visualization} visualizes the CWE distribution across these ten most vulnerable projects.

\begin{table}[htbp]
\centering
\footnotesize
\caption{Summary of vulnerability classification ensemble and analysis results across the 21,430 functions.}
\label{tab:vulnerability_summary}
\begin{tabular}{l@{\hspace{1.5cm}}r}
\hline
\multicolumn{1}{l}{\textbf{Analysis Tools}} &
\multicolumn{1}{r}{Functions} \\
\hline
Semgrep & 2,459 (\textbf{11.5\%}) \\
Flawfinder & 2,091 (\textbf{9.8\%}) \\
Pattern Matching & 1,949 (\textbf{9.1\%}) \\
\hline
\multicolumn{1}{l}{\textbf{Dataset Classification}} &
\multicolumn{1}{r}{Functions} \\
\hline
Total functions & 21,430 \\
Vulnerable & 1,540 (\textbf{7.2\%}) \\
Safe & 17,211 (\textbf{80.3\%}) \\
Uncertain & 2,679 (\textbf{12.5\%}) \\
\hline
\end{tabular}
\end{table}

\begin{table}[t]
\centering
\footnotesize
\caption{Top CWE categories among the 1,540 ensemble-vulnerable functions.}
\label{tab:top_cwe}
\begin{tabular}{lc}
\hline
\textbf{CWE / Vulnerability Type} & \textbf{Vul. Functions} \\
\hline
CWE-120: Buffer Copy & 494 \\
CWE-787: Out-of-Bounds Write & 478 \\
CWE-676: Dangerous Function Use$^{*}$ & 411 \\
CWE-22: Path Traversal & 75 \\
CWE-190: Integer Overflow & 45 \\
CWE-78: OS Command Injection & 25 \\
CWE-134: Format String & 6 \\
CWE-338: Weak PRNG & 6 \\
\hline
\textbf{Total} & \textbf{1,540} \\
\hline
\end{tabular}\\
\footnotesize $^{*}$Dangerous functions includes CWE-369 (Divide by Zero), CWE-476 (Null Pointer Dereference), CWE-362 (Race Condition), CWE-416 (Use After Free), CWE-807 (Untrusted Inputs).
\end{table}

\begin{table}[htbp]
\centering
\footnotesize
\caption{Flawfinder finding distribution by category.}
\label{tab:flawfinder_categories}
\begin{tabular}{lc}
\hline
\textbf{Category} & \textbf{Classified Findings} \\
\hline
Buffer-related    & 1,307 \\
Miscellaneous (e.g. divide by zero)      & 101 \\
Formatted-I/O operations    & 41 \\
Dangerous shell commands     & 30 \\
Race-condition / TOCTOU-related      & 26 \\
Integer-related  & 15 \\
Unsafe temporary-file handling   & 13 \\
Unsuitable random-number generation    & 6 \\
Use of obsolete/deprecated functions  & 6 \\
\hline
\textbf{Total} & \textbf{1,545} \\
\textbf{Unclassified} (but dangerous func.)  & \textbf{546} \\
\hline
\end{tabular}
\end{table}

\begin{table}[t]
\centering
\footnotesize
\caption{Top 10 projects with the highest number of ensemble-vulnerable functions.}
\label{tab:top_projects_cwe}
\begin{tabular}{lcr}
\hline
\textbf{Project} & \textbf{Vuln. Func.} & \textbf{(\%)} \\
\hline
nature-lang/nature & 227 & 14.74\% \\
Serial-Studio/Serial-Studio & 210 & 13.64\% \\
sqliteai/warp & 108 & 7.01\% \\
google/security-research & 77 & 5.00\% \\
xroche/httrack & 72 & 4.68\% \\
ggml-org/whisper.cpp & 50 & 3.25\% \\
memovai/mimiclaw & 50 & 3.25\% \\
DarkFlippers/unleashed-firmware & 37 & 2.40\% \\
DavidXanatos/TaskExplorer & 29 & 1.88\% \\
coturn/coturn & 28 & 1.82\% \\
\hline
\end{tabular}
\end{table}

\subsection{Data Quality and Reproducibility}
\label{subsec:quality}

\noindent\textbf{Deduplication.}
Duplicate function bodies arising when the same AI suggestion appears in multiple commits or forks were identified by exact SHA-256 content hashing and removed. Of the 29,062 raw extracted functions, 7,632 (26.3\%) were duplicates, yielding the final 21,430 unique records.

\noindent\textbf{Repository diversity.} To prevent any single large repository from dominating the dataset, we applied a initial cap of 300 commits per repository to ensure broad repository coverage while maintaining computational tractability. The 21,430 functions originate from \num{226} distinct repositories spanning 1,684 unique commits dated between November 2022 and September 2026.

\noindent\textbf{Contamination awareness.} Functions are associated with their \texttt{commit\_date} to enable researchers to construct temporal splits and avoid data leakage when training models on pre-cutoff data and evaluating on post-cutoff functions---a practice recommended by Ding et al. \cite{primeVul}.

\noindent\textbf{Availability.} LLMVul is publicly available on Zenodo under a
Creative Commons Attribution 4.0 licence (DOI: \texttt{\url{https://doi.org/10.5281/zenodo.22668216}}).
The mining and labeling scripts are released on GitHub at \url{https://github.com/Wahed08/LLMVul-Dataset}. A sample of 100 functions is included in the repository for immediate exploration without downloading the full dataset.

\begin{table*}[htbp]
\centering
\scriptsize
\setlength{\tabcolsep}{2.5pt}
\caption{Comparison of LLMVul with representative human-written vulnerability, LLM-prompted, and AI-assisted code datasets. LLM-prompted datasets contain code generated through controlled research prompts, whereas AI-assisted datasets contain code produced with AI coding assistants during real-world software development and mined from public repositories.}

\label{tab:dataset_comparison}

\begin{tabular*}{\textwidth}{@{\extracolsep{\fill}}*{12}{c}@{}}
\toprule

\textbf{Dataset, Year of Release}
& \textbf{Purpose}
& \textbf{Code Origin}
& \textbf{Lang.}
& \textbf{Repos}
& \textbf{Unit}
& \textbf{Scale}
& \textbf{Vuln. Label}
& \textbf{CWE Label}
& \textbf{Ground-Truth Method}
& \textbf{Prod. Code}
& \textbf{AI Tool Meta}
\\

\midrule

\multicolumn{12}{c}{
\textit{--- \textbf{Human-written vulnerability datasets} ---}
}
\\
\addlinespace[2pt]

Devign \cite{Devign}, 2019
& Vuln. detection
& Human (CVE)
& C/C++
& 2
& Functions
& 27K+
& \cmark
& \xmark
& CVE-linked
& \cmark
& \xmark
\\
\addlinespace[3pt]

BigVul \cite{Bigvul}, 2020
& Vuln. detection
& Human (CVE)
& C/C++
& 348
& Functions
& 265K+
& \cmark
& \cmark
& CVE-linked
& \cmark
& \xmark
\\
\addlinespace[3pt]

ICVul \cite{icvul2025}, 2025
& Vuln. detection
& Human (CVE)
& C/C++
& 807
& Functions
& 15K+
& \cmark
& \cmark
& Automated
& \cmark
& \xmark
\\

PrimeVul \cite{primeVul}, 2025
& Vuln. detection
& Human (CVE)
& C/C++
& 755
& Functions
& 235K+
& \cmark
& \cmark
& Dedup+verif.
& \cmark
& \xmark
\\
\addlinespace[3pt]

\midrule

\multicolumn{12}{c}{
\textit{--- \textbf{LLM prompted code datasets} ---}
}
\\
\addlinespace[2pt]

SecurityEval \cite{siddiq2022securityeval}, 2022
& LLM security eval
& Prompted
& Python
& ---
& Prompts
& 130
& \cmark
& \cmark
& Manual
& \xmark
& \xmark
\\
\addlinespace[3pt]

FormAI \cite{2023formai}, 2023
& LLM vuln. analysis
& Prompted
& C
& ---
& Programs
& 112K
& \cmark
& \xmark
& Formal verif.
& \xmark
& \xmark
\\
\addlinespace[3pt]

CyberSecEval 2 \cite{bhatt2024cyberseceval2}, 2024
& LLM security eval
& Prompted
& Multi
& ---
& Prompts
& 500
& N/A
& N/A
& N/A
& \xmark
& \xmark
\\
\addlinespace[3pt]

CWEval \cite{cweval25}, 2025
& Security + func. eval
& Prompted
& Multi
& ---
& Functions
& 119
& \cmark
& \cmark
& Dynamic
& \xmark
& \xmark
\\
\addlinespace[3pt]

SafeGenBench \cite{safegenbench2025}, 2025
& LLM security eval
& Prompted
& Multi
& ---
& Functions
& 558
& \cmark
& \cmark
& Automated
& \xmark
& \xmark
\\
\addlinespace[3pt]

\midrule
\multicolumn{12}{c}{
\textit{--- \textbf{AI-assisted public code datasets} ---}
}
\\
\addlinespace[2pt]

DevGPT \cite{devGPT}, 2024
& AI-assist. dev. study
& Real-world
& Multi
& ---
& Snippets
& 19K+
& \xmark
& \xmark
& N/A
& \cmark
& Partial
\\
\addlinespace[3pt]

Debt AI Boom \cite{liu2026debtaiboom}, 2026
& Tech. debt study
& Real-world
& Py/JS/TS
& 6,299
& Commits
& 302.6K
& \xmark
& \xmark
& N/A
& \cmark
& \cmark
\\

\addlinespace[3pt]

\textbf{LLMVul (Ours)}
& \textbf{Vuln. detection for LLM code}
& \textbf{Real-world}
& \textbf{C/C++}
& \textbf{1200}
& \textbf{Functions}
& \textbf{21.4K}
& \cmark
& \cmark
& \textbf{3 Tool Ensemble + $\kappa$=0.79}
& \cmark
& \cmark
\\

\bottomrule
\end{tabular*}

\smallskip

\noindent\footnotesize
\textbf{Key:}
\cmark~= available/yes;
\xmark~= not available/no;
Prod.\ Code = mined from real production repositories (not controlled experiments);
AI Tool Meta = records which AI coding assistant generated the code;
Ground-Truth Method = methodology used to establish ground-truth labels;
Py/JS/TS = Python, JavaScript, TypeScript only;
Multi = multiple languages;
--- = not applicable (prompted datasets have no source repository).

\end{table*}

\section{POSSIBLE RESEARCH QUESTIONS}
\label{sec:research-questions}
LLMVul is designed to support empirical studies of the security, generalization, provenance, and evolution of LLM-generated C/C++ code. The dataset supports, among others, the following research questions.

\begin{tcolorbox}[
    colframe=gray,
    colback=white,
    boxsep=1pt,     
    left=1pt,
    right=1pt,
    top=1pt,
    bottom=1pt,
    boxrule=1pt,
]
\small \textbf{\underline{Detection \& Generalization}.}\\
\small \textbf{RQ1}: Do state-of-the-art vulnerability detectors trained on human-written code (e.g., LineVul, VulBERTa, and LLMxCPG) generalize to LLM-generated C/C++ functions, or do they exhibit significant F1 degradation relative to their performance on human-written code?\\
\textbf{RQ2}: Can a vulnerability detection model fine-tuned on LLMVul outperform general-purpose detection models trained on human-written code datasets such as Devign and PrimeVul when evaluated on LLM-generated code?
\end{tcolorbox}

\begin{tcolorbox}[
    colframe=gray,
    colback=white,
    boxsep=1pt,     
    left=1pt,
    right=1pt,
    top=1pt,
    bottom=1pt,
    boxrule=1pt,
]
\small \textbf{\underline{Security Characterization}.}\\
\small \textbf{RQ3}: Does LLM-generated C/C++ code exhibit a distinct CWE distribution from human-written vulnerable code in Devign and PrimeVul, and does this distribution vary across AI coding tools such as Copilot, ChatGPT, Claude, and Cursor?\\
\textbf{RQ4}: How has the proportion of vulnerable functions in LLM-generated C/C++ code evolved over time (2022--2026), and does it vary across AI coding tools and periods of use?
\end{tcolorbox}

\begin{tcolorbox}[
    colframe=gray,
    colback=white,
    boxsep=1pt,     
    left=1pt,
    right=1pt,
    top=1pt,
    bottom=1pt,
    boxrule=1pt,
]
\small \textbf{\underline{Dataset Utility \& Analysis}.}\\
\small \textbf{RQ5}: Can LLMVul's AI-attribution metadata (ai\_tool, signal\_strength, and signal\_type) support function-level classification of LLM-generated versus human-written C/C++ code? \\
\textbf{RQ6}: How effectively do different components of the static-analysis ensemble (Semgrep, Flawfinder, and pattern matching) detect vulnerabilities in LLM-generated C/C++ code, and do their relative effectiveness rankings differ from those observed on human-written code benchmarks?
\end{tcolorbox}
These research questions illustrate the utility of LLMVul beyond benchmark construction. Its real-world repository provenance, vulnerability labels, CWE information, AI-attribution signals, and temporal metadata enable systematic investigation of how LLM-generated code differs from human-written code and how existing security-analysis techniques generalize to this emerging code population.

\section{THREATS TO VALIDITY}
\label{sec:threats}
Several considerations regarding the scope and interpretation of LLMVul should be noted. First, vulnerability labels are generated using a three-tool static-analysis ensemble with a two-of-three majority-vote criterion. Static analysis may produce false positives or false negatives when applied to isolated function fragments. To assess the reliability of the labeling procedure, we manually validated a sample of 100 functions, obtaining ($\kappa = 0.79$). CodeQL was not included because its C/C++ extractor requires compilable translation units, which are incompatible with our function-level representation \cite{codeql2025buildmode}. Second, AI attribution relies on explicit Git metadata signals. Consequently, AI-assisted functions for which developers did not record AI usage in commit metadata may not be identified, potentially causing LLMVul to underestimate the prevalence of AI-generated code. Moreover, substantial post-generation editing may make LLM-generated code less distinguishable from human-written code, which may limit the effectiveness of provenance-based analyses.

Finally, LLMVul samples the top 1,200 C/C++ repositories by star count to focus on widely used, actively maintained software projects. This sampling strategy favors mature and widely adopted projects and therefore may not fully represent smaller, private, less-established, or embedded codebases. LLMVul currently focuses on C and C++; whether the observed patterns generalize to other programming languages, such as Python and JavaScript, remains an open question.

\section{CONCLUSION \& FUTURE WORK}
\label{sec:conclusion}
We introduce LLMVul, a vulnerability dataset of C/C++ functions mined from authentic GitHub repositories containing code developed with AI coding assistants. Unlike controlled or simulated benchmarks, LLMVul captures vulnerabilities that arise incidentally during real-world AI-assisted development. The dataset comprises thousands of functions and commits, with vulnerabilities spanning multiple CWE categories and labels derived through a static-analysis ensemble and validated by human raters. Each function is accompanied by commit-level provenance and AI-tool attribution, enabling systematic investigation of the security, provenance, and generalization of LLM-generated code.

As future work, we plan to expand LLMVul to additional programming languages, including Python, Java, and Rust, enabling cross-language analysis of vulnerabilities in AI-generated code. We also plan to develop automated build infrastructure that supports additional static analyzers and enables analysis of compilable code contexts, with the goal of improving vulnerability detection and memory-safety characterization.

\section*{ACKNOWLEDGEMENT}
This research was funded in part by U.S. National Science Foundation under Grant Number OIA-2437963 and Louisiana Board of Regents.

\bibliographystyle{ACM-Reference-Format}
\bibliography{sample-new}

\end{document}